\documentclass{aastex63}
\usepackage{CJK}
\usepackage{amsmath}

\usepackage[T1]{fontenc}

\shorttitle{SMARPs and SHARPs}
\shortauthors{Bobra et al.}
\begin{document}
\begin{CJK*}{UTF8}{gbsn}

\title{SMARPs and SHARPs: Two Solar Cycles of Active Region Data}

\correspondingauthor{Monica G. Bobra}
\email{mbobra@stanford.edu}

\author[0000-0002-5662-9604]{Monica G. Bobra}
\affiliation{W.W. Hansen Experimental Physics Laboratory, Stanford University, Stanford, CA 94305, USA}

\author[0000-0001-9021-611X]{Paul J. Wright}
\affiliation{W.W. Hansen Experimental Physics Laboratory, Stanford University, Stanford, CA 94305, USA}

\author[0000-0003-4043-616X]{Xudong Sun (孙旭东)}
\affiliation{Institute for Astronomy, University of Hawai`i at M\={a}noa, Pukalani, HI 96768-8288, USA}

\author[0000-0002-6463-063X]{Michael J. Turmon}
\affiliation{Jet Propulsion Laboratory, California Institute of Technology, Pasadena, CA 91109, USA}

\begin{abstract}
We present a new data product, called Space-Weather MDI Active Region Patches (SMARPs), derived from maps of the solar surface magnetic field taken by the Michelson Doppler Imager (MDI) aboard the {\it Solar and Heliospheric Observatory} (SoHO). Together with the Space-Weather HMI Active Region Patches (SHARPs), derived from similar maps taken by the Helioseismic and Magnetic Imager (HMI) aboard the {\it Solar Dynamics Observatory}, these data provide a continuous and seamless set of maps and keywords that describe every active region observed over the last two solar cycles, from 1996 to the present day. In this paper, we describe the SMARP data and compare it to the SHARP data.

\end{abstract}

\keywords{solar active regions --- astronomy databases --- solar magnetic fields}

\section{Introduction}\label{sec:intro}
We present a new data product, called Space-Weather MDI Active Region Patches (SMARPs), derived from maps of the solar surface magnetic field taken by the Michelson Doppler Imager \citep[MDI,][]{scherrer95} aboard the {\it Solar and Heliospheric Observatory} (SoHO). These data include maps that track every solar active region observed by MDI, along with keywords that describe the physical characteristics of each active region, from 1996 to 2010.

We designed the SMARP data for use in concert with another data product called the Space-Weather HMI Active Region Patches \citep[SHARPs,][]{bobra14}, derived from photospheric magnetic field data taken by the Helioseismic and Magnetic Imager \citep[HMI,][]{schou12} aboard the {\it Solar Dynamics Observatory} (SDO). The SHARP data include tracked active region maps and summary parameters from 2010 to the present day. These data motivated several studies, such as quantifying electric currents \citep[e.g.][]{Kontogiannis2017} and flux cancellation \citep[e.g.][]{Yardley2016} within solar active regions, connecting photospheric magnetic field properties to kinematic properties of coronal mass ejections \citep[e.g.][]{murray2018}, developing high-resolution magnetohydrodynamic simulations of active regions as they evolve over time \citep[e.g.][]{hayashi2018}, and forecasting solar flares using the keyword metadata \citep[e.g.][]{chen2019} and image data \citep[e.g.][]{Deshmukh2020}.

However, HMI observations of the solar photosphere coincide with Solar Cycle 24, the weakest solar cycle in a century \citep{sidc}. MDI, on the other hand, took observations during the far stronger Solar Cycle 23. Combined, the SMARP and SHARP databases provide a continuous, seamless set of active region data from both Cycle 23 and 24. Together, these data provide an opportunity to extend statistical studies to a larger sample of strong, complex active regions. In this paper, we describe the SMARP data and compare it to the SHARP data.

Users can access SMARP data with an open-source and openly developed Python package for solar physics data analysis called SunPy \citep{sunpy_community2020}, which uses an affiliated package called drms \citep{Glogowski2019drms} to access the Joint Science Operations Center (JSOC; \href{http://jsoc.stanford.edu/}{jsoc.stanford.edu}). JSOC serves data from MDI, HMI, and other solar instruments. Alternatively, users can access SMARP data from the JSOC web interface. We also provide example code that demonstrates how to use the SMARP and SHARP data together (see Table \ref{tab:drmsseries}).

\section{Methodology}\label{sec:data}

\begin{figure*}
\gridline{\fig{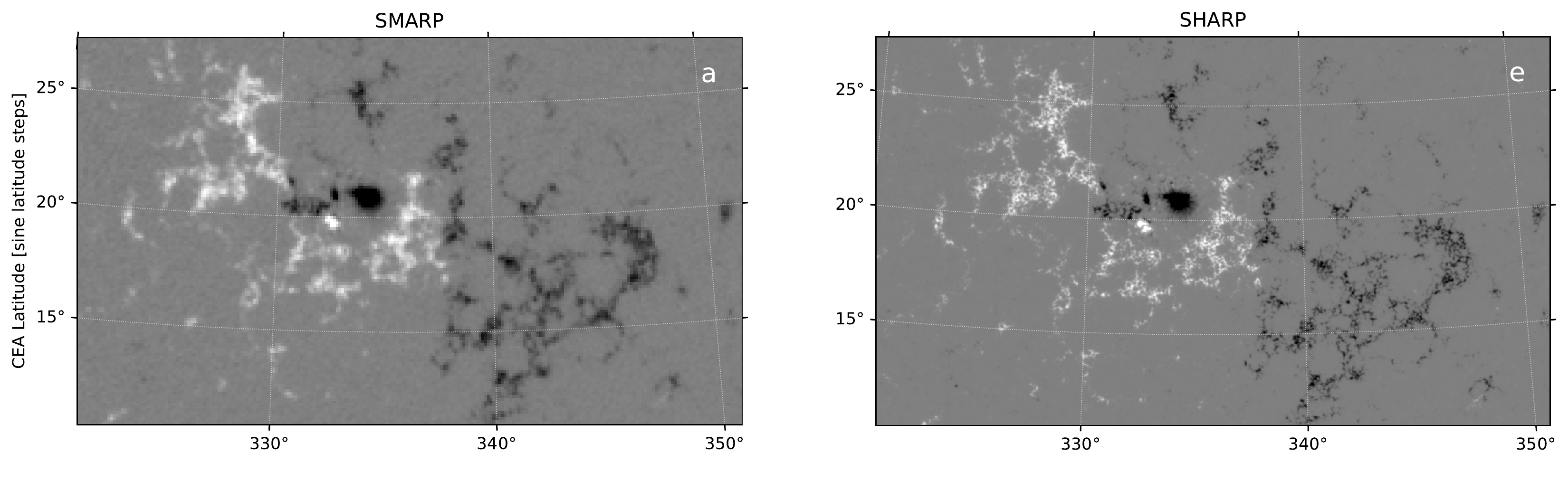}{0.95\textwidth}{}}
\vspace{-16mm}
\gridline{\fig{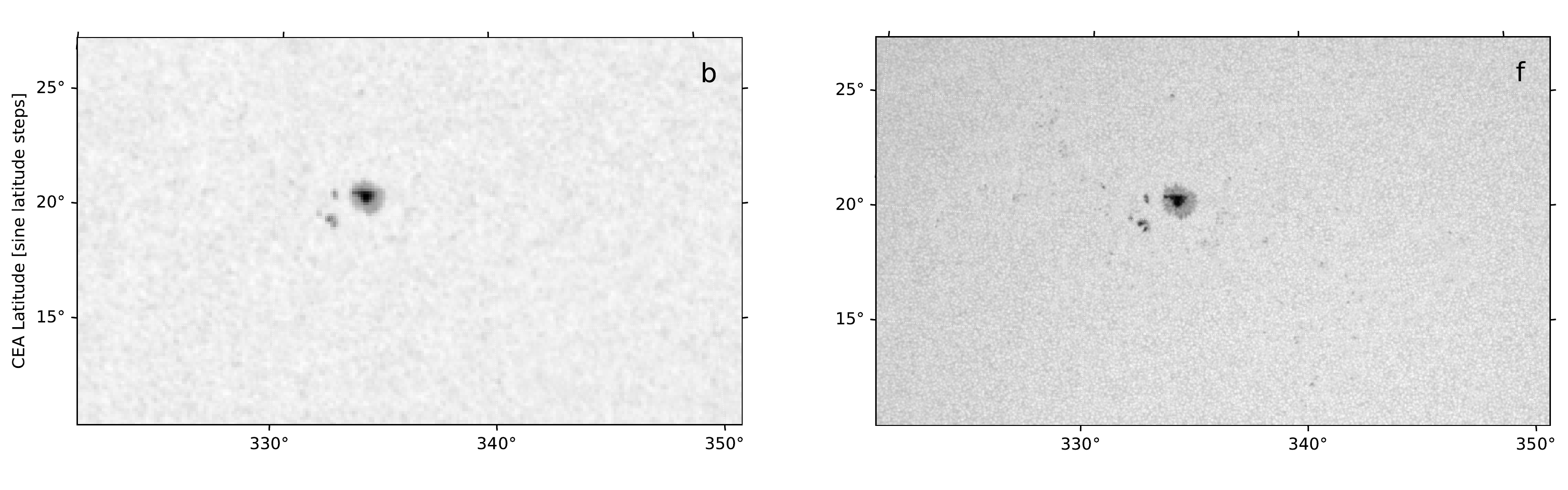}{0.95\textwidth}{}}
\vspace{-16mm}
\gridline{\fig{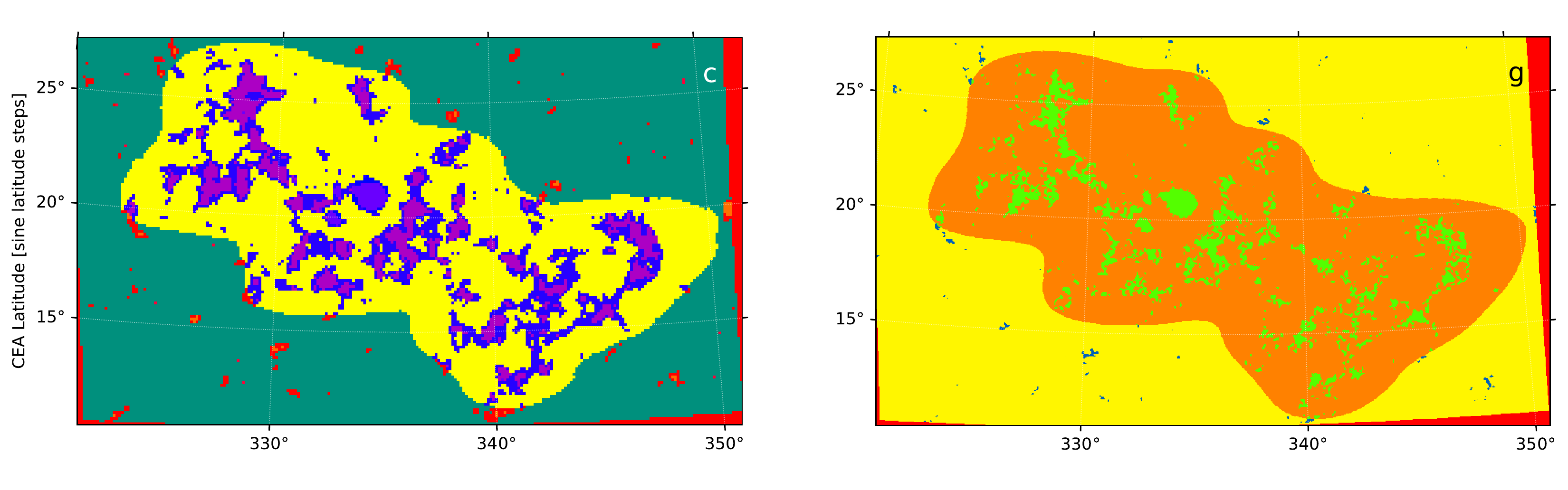}{0.95\textwidth}{}}
\vspace{-18mm}
\gridline{\fig{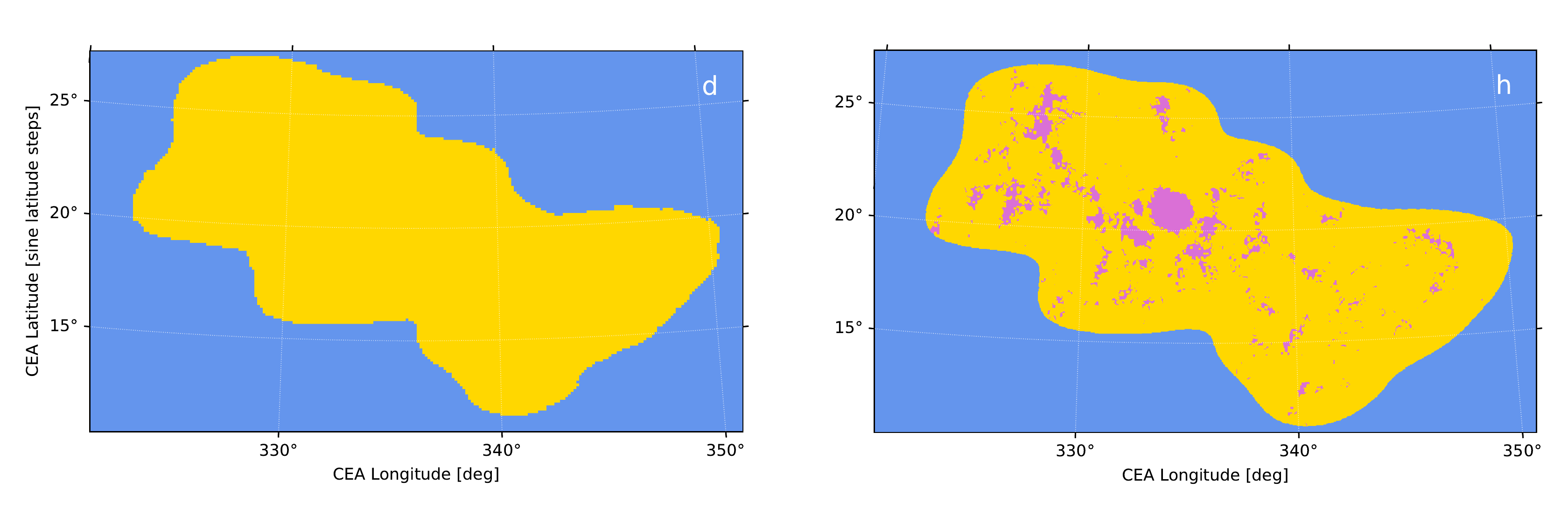}{0.95\textwidth}{}}
\vspace{-10mm}
\caption{An example of the active region detection algorithm for the SMARP (left column) and SHARP (right column) data is shown in Cylindrical Equal-Area coordinates. Both encapsulate NOAA Active Region 11087, which corresponds to TARP \#13520 and HARP \#86, on 14 July 2010 at 11:12:00 TAI. The SMARP detection algorithm identified a bounding box that spans 16.92$^\circ$ in CEA Latitude and 29.04$^\circ$ in CEA Longitude; the SHARP bounding box measures 16.98$^\circ \times$ 29.43$^\circ$. The SMARP and SHARP center coordinates are at (336.17$^\circ$, 19.42$^\circ$) and (335.82$^\circ$, 19.53$^\circ$). The top row shows photospheric line-of-sight magnetic field maps, clipped to $\pm$ 800 Gauss, within the SMARP and SHARP bounding boxes. The second and third rows show the continuum intensity maps and activity bitmaps. The bottom row shows a simplified version of the activity bitmaps, highlighting the pixels that contribute to the value of any given active region summary parameter (see Table \ref{tab:keys}). The gold region in Panel d shows all the pixels in the TARP region, or the pixels in Panel c with values $>$36. The gold and violet region in Panel h shows all the pixels in the HARP region, or the pixels in Panel g with values $>$30. The violet region in Panel h labels pixels with a high signal-to-noise ratio in the transverse component of the photospheric magnetic field. 
\label{fig:Figure1}}
\end{figure*}

The SMARP data provide a complete record of all the active regions observed by MDI between January 1996 and October 2010. The data describe each active region at a 96-minute cadence throughout its entire lifetime, starting two days before it emerges or rotates onto the solar disk until two days after it submerges or disappears from view behind the limb. Specifically, the SMARP data product includes two components: metadata keywords, including 3 that describe the physical characteristics of solar active regions, and 3 partial-disk maps that encompass these regions --- a photospheric line-of-sight magnetic field map, a continuum intensity map, and an activity bitmap (see Figure \ref{fig:Figure1}a-c). 

Unlike its successor, HMI, which retrieves all three components of the magnetic field vector at the solar surface, MDI only retrieved the line-of-sight component of the surface magnetic field. These MDI full-disk line-of-sight maps span $1024\times1024$ pixels, with a pixel size of 2 arcseconds \citep{scherrer95}. The MDI instrument creates two different types of magnetic field maps: a {\it one-minute} magnetic field map, created from filtergrams obtained over a 30-second interval, and a {\it five-minute} magnetic field map (or an average of five one-minute maps). Typically, the MDI instrument produces one of these maps at a 96-minute cadence. The MDI instrument team collated both the one-minute and five-minute magnetic field maps into a singular data series with a cadence of 96 minutes. \citet{liu2012} determined a median noise level of 26.4 Mx cm$^{-2}$ for the one-minute magnetic field maps, and 16.2 Mx cm$^{-2}$ for the five-minute maps (see their Figure 9). The MDI instrument took data from December 19, 1995 to April 11, 2011. See the MDI Resident Archive at \href{http://soi.stanford.edu/mdi}{soi.stanford.edu/mdi} for an overview of all the MDI data products, which are publicly available through the JSOC. For a detailed description of each observable, see
\href{http://soi.stanford.edu/mdi/observables}{soi.stanford.edu/mdi/observables}. 

The steps below describe the SMARP data pipeline. This pipeline includes multiple data series, or collections of metadata keywords and image data. Users can access all of these data series, listed in Table \ref{tab:drmsseries}, via JSOC.

\begin{enumerate}
\item{The MDI Resident Archive project migrated all of the data taken by MDI into JSOC. This involved a number of tasks, including the population of WCS-compliant ephemeris keywords and a final correction of the instrument plate scale. See the data series \texttt{mdi.scale\_corrections} for the final scale corrections table. For a detailed description of the 96-minute magnetic field maps, including the release notes, see \href{http://soi.stanford.edu/mdi/observables/fd_m_96m_lev182.html}{soi.stanford.edu/mdi/observables/fd\_m\_96m\_lev182.html}.}
\item{We created interpolated continuum intensity data  co-temporal with the line-of-sight magnetic field data by de-rotating pairs of flattened continuum intensity maps that bracket a line-of-sight magnetic field map and merging these bracketing pairs via a weighted average. This leaves 79,515 pairs of co-temporal continuum intensity and line-of-sight magnetic field maps at a 96-minute cadence. If the time between any bracketing pair of continuum intensity maps exceeds 36 hours, the temporal interpolation performs poorly and we flag these data by setting a bit in the {\tt QUALITY} keyword. Since these gaps became more frequent near the end of the MDI mission, we do not include data beyond October 28, 2010.\footnote{For more documentation about the temporal interpolation, see
\href{http://jsoc.stanford.edu/data/mdi/mdi-tarp/mdi-interp-ic}{jsoc.stanford.edu/data/mdi/mdi-tarp/mdi-interp-ic}.}} 

\item{For each line-of-sight magnetic field map within this set of 79,515 pairs of co-temporal continuum intensity and line-of-sight magnetic field maps, we compute a full-disk magnetic activity bitmap that identifies active pixels using a Bayesian approach described in \citet{turmon2002, turmon2010}. The core idea is that we optimize a function that balances the per-pixel observations with the local structure. This function contains two terms. The first term, which is dominant in the calculation, matches observed magnetic field values to the expected scatter for each class (i.e. quiet-Sun and active region). The second term, or smoothness term, only affects bitmap values values that are near the boundary between classes. Geometrical information is encoded in the smoothness term. The overall effect does not correspond to a threshold rule on the magnetic field data. Boundaries between classes follow non-axis-parallel contours that are a property of the competition between the expected scatter of the two classes. In sum, active pixels identify spatially coherent regions of strong magnetic field. This is the same identification approach used to determine active pixels in the HMI bitmaps; we simply rescaled the HMI region-identification parameters to account for the differences between the HMI and MDI magnetic field maps as described in \citet{liu2012}. Separately from this, we compute an ancillary three-class bitmap that labels sunspots, faculae, and quiet Sun regions \citep{turmon2002}. We do not use this three-class continuum intensity bitmap for tracking, but we fold the bit fields into the final product.}
\item{We identified active regions by convolving the full-disk magnetic field activity bitmaps with a template (in this case, an elongated Gaussian kernel of FWHM $\approx 50\times25$~Mm, or $\approx 40\times20$~MDI pixels, at disk center). This method, known as a matched filter \citep{turin60}, identifies groups of active pixels on the scale of a typical NOAA active region. We call each group of pixels an active region. We aim to identify all the coherent regions of magnetic activity observed in line-of-sight magnetic field maps, which includes regions that span a small heliographic area and regions without an associated photometric sunspot, since these can can trigger solar flares \citep[e.g.][]{Guennou2017} and disrupt the global magnetic structure \citep[e.g.][]{schrijver2013}.}
\item{We used these grouped full-disk activity bitmaps to track active regions over time. To do this, we extrapolated a bitmap corresponding to $t_{n-1}$ forward one time step, to $t_{n}$, using standard latitude-dependent solar rotation rates. Then we computed an overlap score for each group of pixels in the motion-compensated bitmap and the true bitmap at $t_{n}$. We used these overlap scores to determine the likely movement of each group of pixels between successive bitmaps. We repeated this process for all $t$ to track each group of pixels, or active region, over time.}
\item{We extracted partial-disk patches from the grouped full-disk activity bitmaps. In addition, we fold the three-class continuum intensity bitmap values into these partial-disk patches. These partial-disk activity bitmaps encapsulate the maximum heliographic extent of any given active region. We assigned each region its own identification number, or Tracked Active Region Patch (TARP) Number. For example, Figure \ref{fig:Figure1}c shows TARP 13520, or NOAA Active Region 11087, outlined in yellow.\footnote{See \href{http://jsoc.stanford.edu/data/mdi/mdi-tarp/mdi-browse}{jsoc.stanford.edu/data/mdi/mdi-tarp/mdi-browse} to view time evolution movies of these activity bitmaps.} While we assign a new TARPNUM per active region disk passage, users can link consecutive disk passages of any given active region with the metadata keyword \texttt{NOAA\_ARS}. The partial-disk activity bitmaps label magnetic field strength (quiet or active, encoded with a 1 or 2, respectively), photospheric features (quiet Sun, faculae, or sunspots, encoded with a 4, 8, and 16), and pixel locations (off-disk or a member of the patch, encoded with a 0 or 32). As a result, each pixel can take one of 13 possible values.}
\item{We also extracted partial-disk maps from the full-disk interpolated continuum intensity data and full-disk line-of-sight magnetic field data. Together with the partial-disk bitmaps, this yields three partial-disk maps. We provide these three maps in two coordinate systems that adhere to the World Coordinate System \citep[WCS,][]{thompson2006} standard: Helioprojective Cartesian and Heliographic Cylindrical Equal-Area. The latter places the origin at the center of the patch \citep[for more information, see][]{sun2013}.}
\item{Finally, we calculated three active region summary parameters: the total line-of-sight unsigned flux, mean line-of-sight field gradient, and $R$, a quantity that parameterizes the unsigned flux near polarity inversion lines \citep{schrijver07}. These parameters, which we provide as metadata keywords, describe the physical characteristics of every active region at a 96-minute cadence. We describe these parameters in detail in Section \ref{subsec:keys}.}
\end{enumerate}

\begin{table}
\centering
\caption{SMARP Data and Code}
\begin{tabular}{p{0.25\textwidth}p{0.65\textwidth}}
\hline
\hline
Data Series Name & Description \\
\hline
\texttt{mdi.fd\_M\_96m\_lev182} & Full-disk photospheric line-of-sight magnetic field maps \\
\texttt{mdi.fd\_Ic\_interp} & Full-disk interpolated continuum intensity maps co-temporal with the photospheric line-of-sight magnetic field maps\\
\texttt{mdi.fd\_Marmask} & Full-disk activity bitmaps that label regions of strong and weak magnetic field \\
\texttt{mdi.fd\_spotmask} & Full-disk activity bitmaps that label sunspots, faculae, and quiet Sun regions \\
\texttt{mdi.smarp\_96m} & Partial-disk magnetic and photospheric activity bitmaps, line-of-sight magnetic field maps, and continuum intensity maps in Helioprojective Cartesian coordinates; space-weather metadata keywords \\
\texttt{mdi.smarp\_cea\_96m} & Partial-disk magnetic and photospheric activity bitmaps, line-of-sight magnetic field maps, and continuum intensity maps in Heliographic Cylindrical Equal-Area coordinates; space-weather metadata keywords \\
\hline
\hline
Code Repository & Description \\
\hline
\href{http://jsoc.stanford.edu/cvs/JSOC/proj/sharp/apps/}{jsoc.stanford.edu/cvs/JSOC} & This repository contains all the code to run the JSOC software pipeline. See the SMARP pipeline code (smarp.c and smarp\_functions.c) and SHARP pipeline code (sharp.c and sw\_functions.c) under /proj/sharp/apps. \\
\href{https://github.com/mbobra/SHARPs}{github.com/mbobra/SHARPs} & This repository contains code that demonstrates how to use the SHARP data. This repository is published in Zenodo as v0.1.0 and citeable with the DOI 10.5281/zenodo.5131292 \citep{monica_g_bobra_sharp}.\\
\href{https://github.com/mbobra/SMARPs}{github.com/mbobra/SMARPs} & This repository contains code that demonstrates how to use the SHARP data. This repository is published in Zenodo as v0.1.0 and citeable with the DOI 10.5281/zenodo.5138025 \citep{monica_g_bobra_smarp}.\\
\end{tabular}
\label{tab:drmsseries}
\end{table}

\begin{figure*}
\gridline{\fig{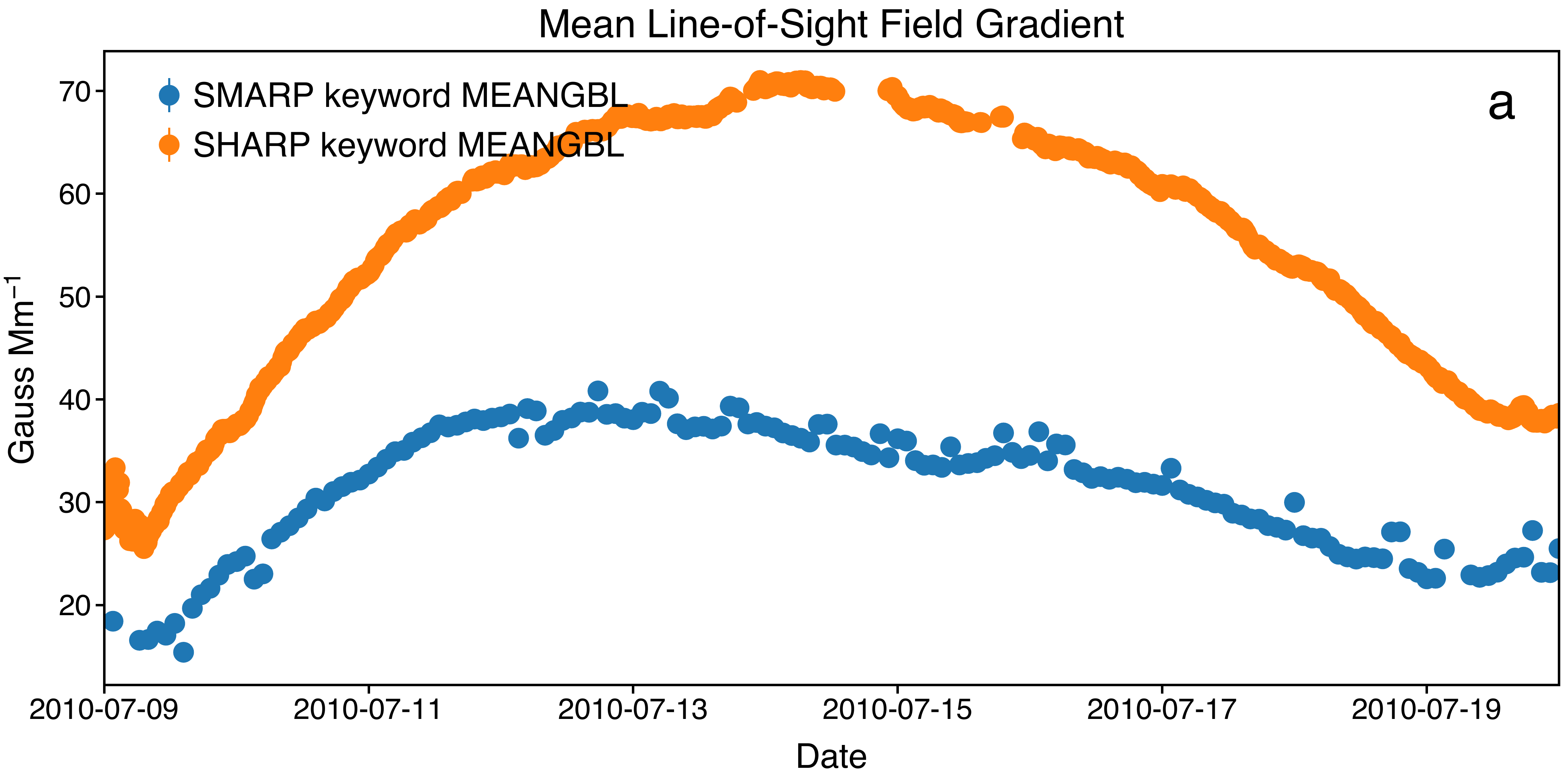}{0.495\textwidth}{}
          \fig{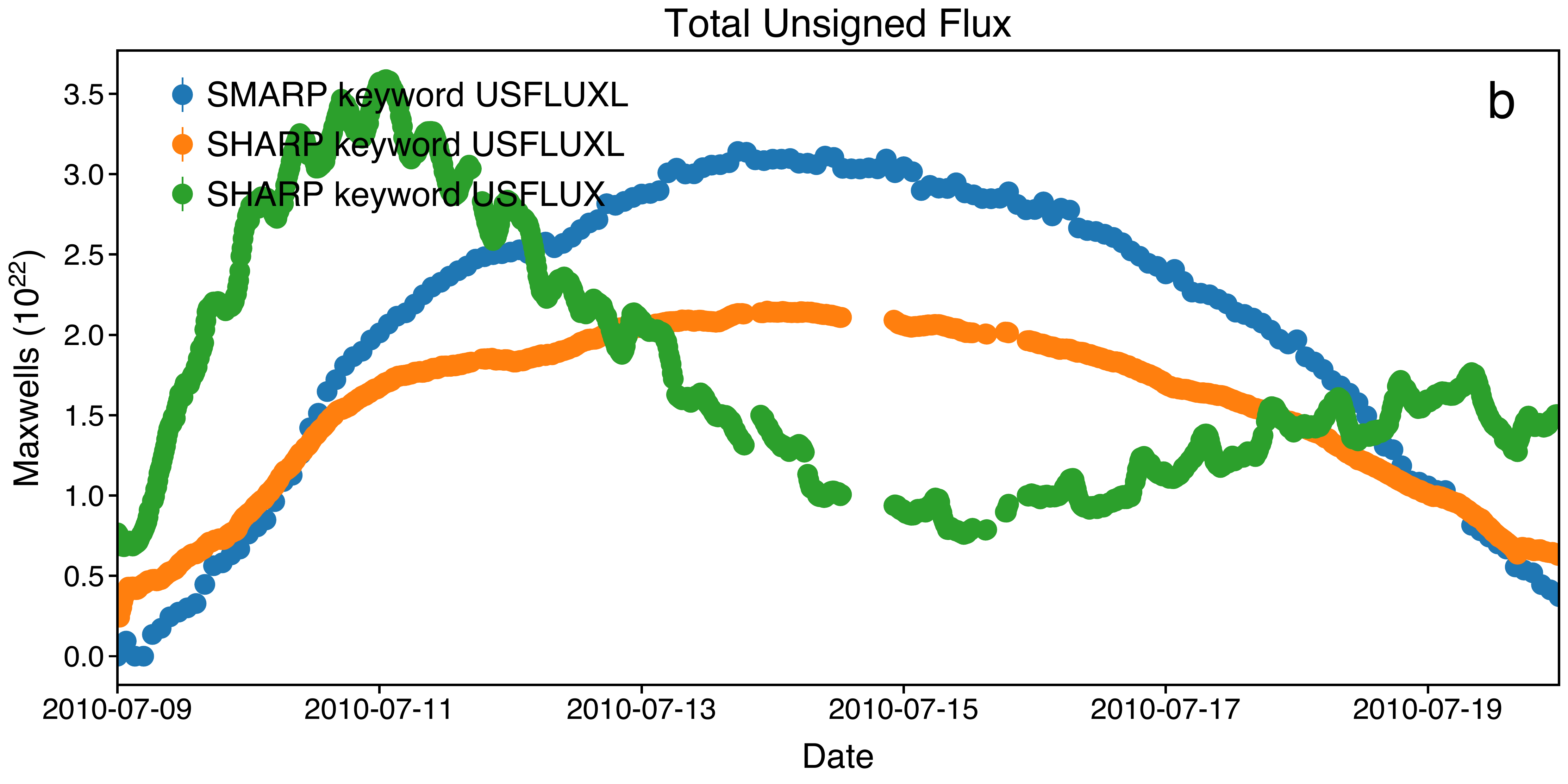}{0.495\textwidth}{}}
\vspace{-8mm}
\gridline{\fig{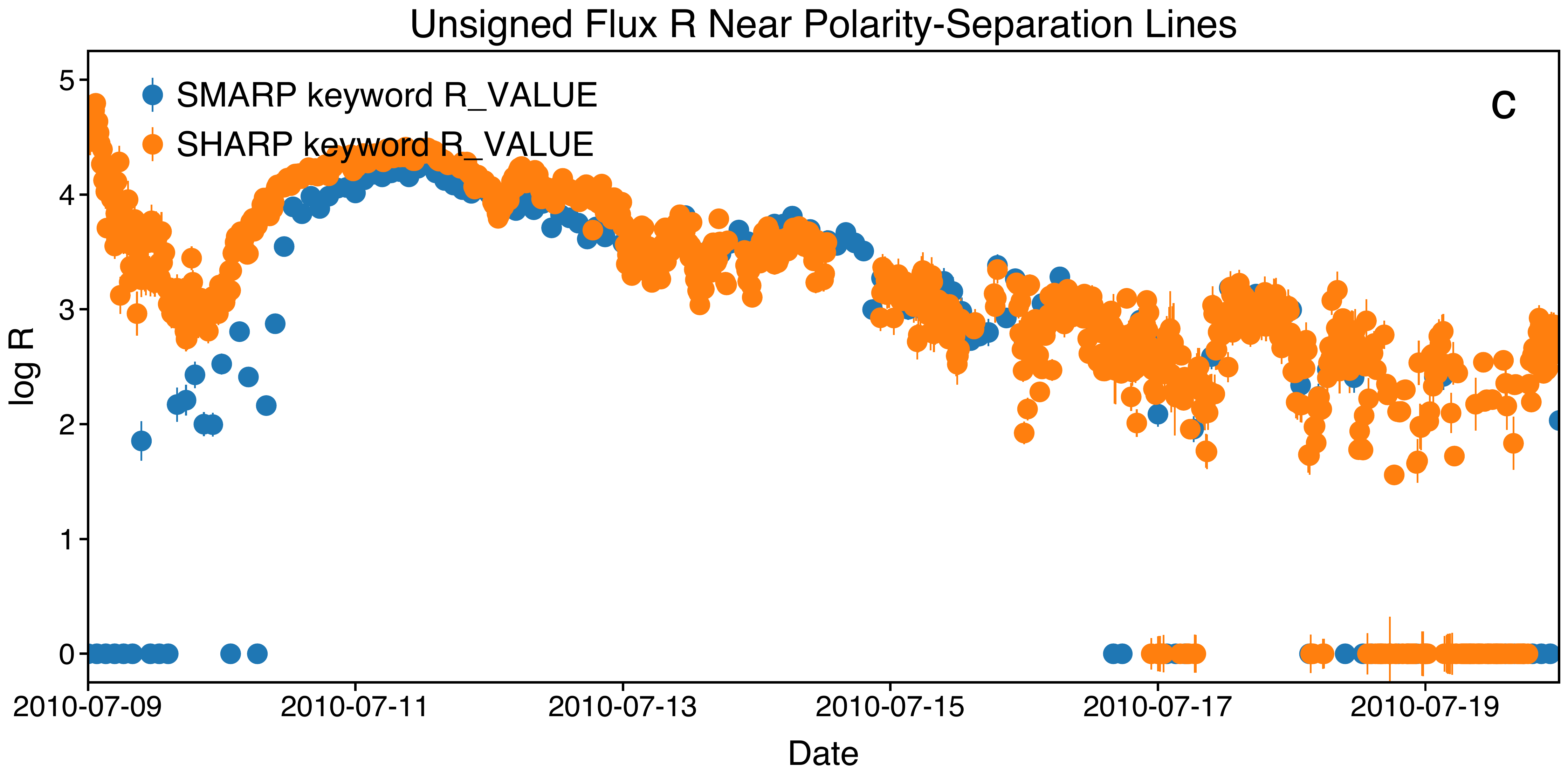}{0.495\textwidth}{}
          \fig{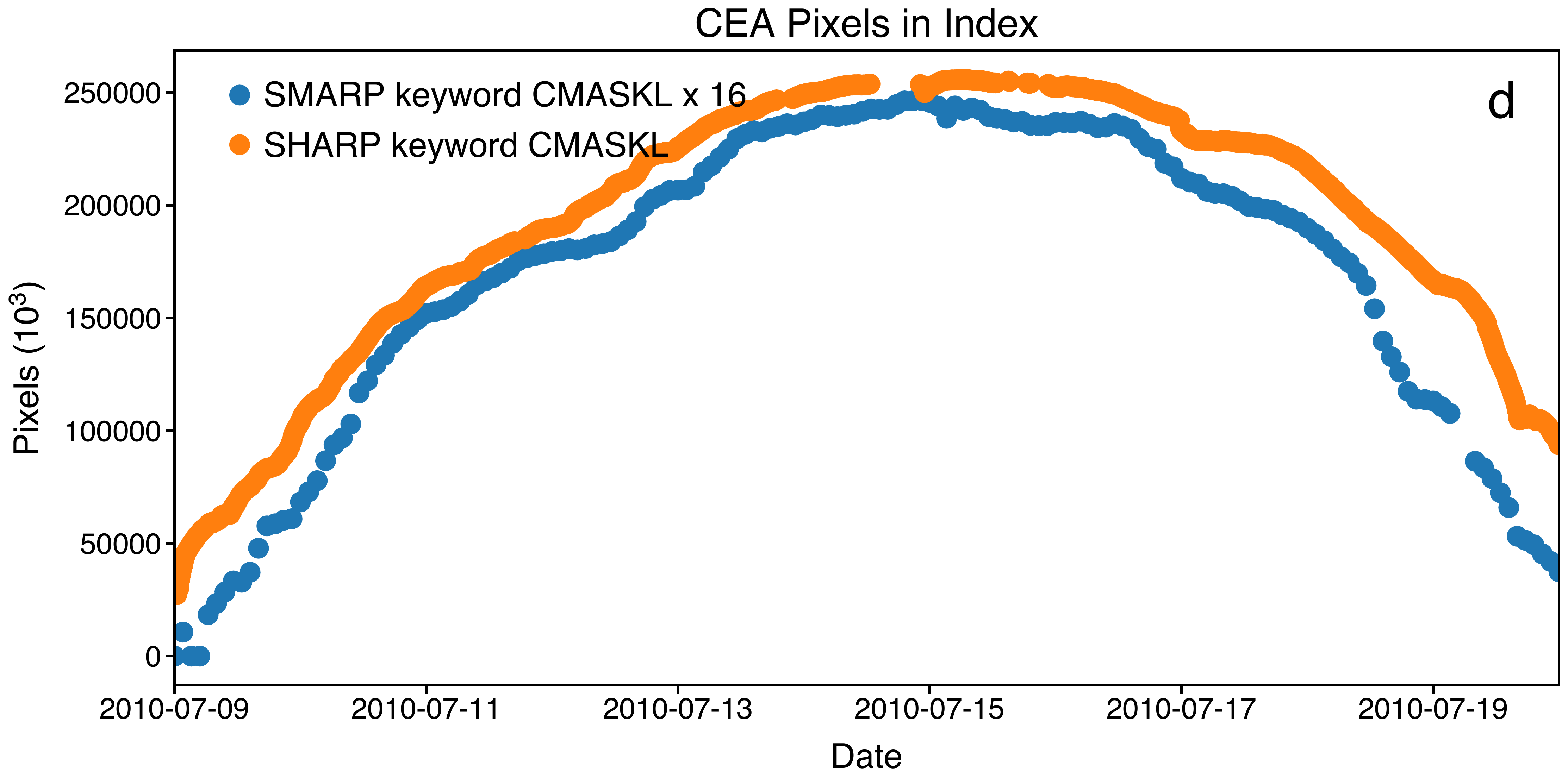}{0.495\textwidth}{}}
\vspace{-8mm}
\caption{This figure shows SMARP (blue) and SHARP (orange) summary parameters, derived from the line-of-sight magnetic field data, for NOAA Active Region 11087 from 9 July 2010 to 20 July 2010. The first three panels show the mean line-of-sight field gradient, total line-of-sight unsigned flux, and $R$, or the logarithm of the total unsigned flux near polarity inversion lines. We calculate these parameters using the methods described in Table \ref{tab:keys}. Panel b also shows the total unsigned flux in green, derived from the vertical component of the vector magnetic field data, for the same active region during the same time. The oscillations in this quantity arises from a Doppler shift in the HMI spectral line due to the orbital variation of the SDO spacecraft \citep[see Section 3.2 of][]{couvidat2016}. See Table 3 and Panel A4 of Figure 4 in \citet{bobra14} for more information. Note that all the quantities plotted in panels (a) - (c) include error bars, but most of the error bars are smaller than the size of the dots. Panel d indicates the number of pixels that contributed to the value of any given active region summary parameter. To account for the difference in spatial resolution between the SMARP and SHARP image data, we multiplied the SMARP \texttt{CMASKL} values by the square of the SMARP to SHARP plate scale ratio for image data projected in the Cylindrical Equal-Area coordinate system. This equals ${\left(\frac{0.12^{\circ}/\text{pixel}}{0.03^{\circ}/\text{pixel}}\right)}^2$ or 16. Note that the degree to which the line-of-sight field deviates from the radial field affects the time variation of any given summary parameter. See Section \ref{subsec:limits} for more information about limitations of the SMARP data.
\label{fig:Figure2}}
\end{figure*}

\section{Comparing SMARP and SHARP Data: The Overlap Period}\label{sec:overlap}
\subsection{Active Region Maps}\label{subsec:maps}

The SMARP database contains 6063 TARPs tracked throughout their lifetime. About half, or 2975, of these regions are associated with one or more NOAA active region numbers. Our definition of an active region, described in Step 4 of Section \ref{sec:data}, differs from that of NOAA's. 

As described in \citet{Giersch2018} and \citet{meadows2020}, NOAA manually identifies active regions from hand-drawings of white-light observations created by the United States Air Force Solar Observing Optical Network (SOON), who use a sunspot-area overlay mask \citep[see Figure 4 of][]{Giersch2018} to determine the size of any given region. We automatically identify active regions in line-of-sight magnetic field maps using the software algorithm described in Section \ref{sec:data}. We count regions that span a small heliographic area and regions without an associated photometric sunspot as active regions. As a result, both the SMARP and SHARP data sets include more regions than the NOAA active region database. 

The SMARP and SHARP data overlap for half a year, between 1 May 2010 and 28 October 2010. This overlap period provides an opportunity to characterize the differences between the two data sets. During this period, NOAA identified 57 active regions. The SMARP and SHARP data sets contain 127 and 137 active regions, respectively, during the same time period.

Two factors contribute to the discrepancy between the number of active regions in the SMARP and SHARP data sets during the overlap period. First, MDI did not observe continuously throughout the overlap period. HMI, however, observes the Sun nearly continuously \citep{schou12}. Second, differences in target line, and, most significantly, spatial resolution between the MDI and HMI instruments \citep[see Table 2 of][]{schou12} affect the inferred magnetic field values for any given pixel \citep{liu2012}. The ten discrepant active regions that appear in the SHARP, but not in the SMARP, database span an area of 860 micro-Hemispheres or less; furthermore, NOAA did not label any of these regions with an active region number.

These factors also affect active regions common to both data sets. Figure \ref{fig:Figure1} shows an example. Even though we use the same method to detect active regions in the SMARP and SHARP data, NOAA Active Region 11087 differs slightly in shape, size, and position. Furthermore, the MDI line-of-sight magnetic field data include some optical distortion with a temporal and spatial variance that is, as of now, not fully understood. \citet{liu2012} note that the optical distortion in the MDI data arise from instrumental imperfections in the left- and right-circular analyzers, which enable signals from granulation and $p$-mode oscillations to leak in as noise. ``Because the filtergrams are combined onboard,'' they write, ``the distortions cannot be fixed in the MDI data.'' As a result, a coordinate in the SHARP data may not map to the same physical feature in the SMARP data. This effect is almost zero at disk-center and up to a few MDI pixels at the limb. 

\subsection{Active Region Summary Parameters}\label{subsec:keys}

\begin{table}
\centering
\caption{Active Region Summary Parameters}
\begin{tabular}{p{0.07\textwidth}p{0.265\textwidth}p{0.372\textwidth}p{0.22\textwidth}}
\hline
\hline
Keyword & Description & Pixels & Formula \\
\hline
\hline
\multicolumn{4}{c}{SMARP Data Series (\texttt{mdi.smarp\_96m} and \texttt{mdi.smarp\_cea\_96m})} \\
\hline
\texttt{USFLUXL} & Total line-of-sight unsigned flux & Pixels in the TARP region (e.g. gold region Figure \ref{fig:Figure1}d) & $\sum|B_{LoS}|dA$ \\
\texttt{MEANGBL} & Mean gradient of the line-of-sight field & Pixels in the TARP region & $\sqrt{\left(\frac{\partial B_{LoS}}{\partial x}\right)^2 + \left(\frac{\partial B_{LoS}}{\partial y}\right)^2}$ \\
\texttt{R\_VALUE} & $R$, or a measure of the unsigned flux near polarity inversion lines & Pixels near polarity inversion lines & $\log\left(\sum|B_{LoS}|dA\right)$ \\
\hline
\multicolumn{4}{c}{SHARP Data Series (\texttt{hmi.sharp\_720s} and \texttt{hmi.sharp\_cea\_720s})} \\
\hline
\texttt{USFLUXL} & Total line-of-sight unsigned flux & Pixels in the HARP region (e.g. gold and violet region Figure \ref{fig:Figure1}h) & $\sum|B_{LoS}|dA$ \\
\texttt{MEANGBL} & Mean gradient of the line-of-sight field & Pixels in the HARP region & $\sqrt{\left(\frac{\partial B_{LoS}}{\partial x}\right)^2 + \left(\frac{\partial B_{LoS}}{\partial y}\right)^2}$ \\
\texttt{R\_VALUE} & $R$ & Pixels near polarity inversion lines & $\log\left(\sum|B_{LoS}|dA\right)$ \\
\texttt{USFLUX} & Total vertical unsigned flux & Pixels in the HARP region and above the high-confidence disambiguation threshold (e.g. violet region in Figure \ref{fig:Figure1}h) & $\sum|B_{z}|dA$ \\
\texttt{MEANGBZ} & Mean gradient of the vertical field & Pixels in the HARP region and above the high-confidence disambiguation threshold & $\sqrt{\left(\frac{\partial B_{z}}{\partial x}\right)^2 + \left(\frac{\partial B_{z}}{\partial y}\right)^2}$ \\
\hline
\hline
\end{tabular}
\label{tab:keys}
\end{table}

The SMARP database contains three active region summary parameters, listed as the first three parameters in Table \ref{tab:keys} under the subheading ``SMARP Data Series'', which describe the physical characteristics of every active region observed by MDI at a 96-minute cadence. We derive these parameters from maps of the line-of-sight magnetic field in Heliographic Cylindrical Equal-Area coordinates. In this projection, the line-of-sight field equals the vertical field at patch center. While the deviation between the line-of-sight and vertical field increases with radial distance from the patch center, it amounts to a few degrees across a typical active region \citep[see Equation 9 of][]{sun2013}. We select pixels in the TARP region (e.g. the gold region in Figure \ref{fig:Figure1}d), to calculate the first two parameters listed in Table \ref{tab:keys}. 

To calculate $R$, we reproduce the method outlined in  \cite{schrijver07}. To identify a polarity inversion line, \citet{schrijver07} [1] create a bitmap by labeling pixels $>$150 Gauss with a 1 and the remaining pixels with 0, [2] create another bitmap by labeling pixels $<$-150 Gauss with a 1 and the remaining pixels with 0, [3] convolve both these bitmaps with a boxcar filter, and [4] create yet another bitmap by identifying the pixels for which both convolved bitmaps equal 1. This bitmap, which we will call $p1$, identifies the polarity inversion lines. To calculate $R$, \citet{schrijver07} [1] convolve $p1$ with a Gaussian filter to identify pixels near the polarity inversion line, [2] multiply the resultant array with the absolute value of the line-of-sight magnetic field map, and [3] sum all the values in the multiplied array, and [4] take the common logarithm of the resultant value. For more detail, see the publicly available code (linked in Table \ref{tab:drmsseries}) and Sections 2 and 3 of \citet{schrijver07}.

The SHARP database originally contained 17 active region summary parameters, which describe the physical characteristics of every active region observed by HMI at a 12-minute cadence. \cite{bobra14} calculated these parameters with an alternate method, by selecting pixels within the HARP region that contain a high signal-to-noise ratio in the transverse field (e.g. the violet region in Figure \ref{fig:Figure1}h) from vector magnetic field maps in Heliographic Cylindrical Equal-Area coordinates. 

As a result, the SMARP and SHARP active region summary parameters show different absolute values and trending behavior. For example, Figure \ref{fig:Figure2}b shows the total unsigned flux over time for NOAA Active Region 11087 as calculated from the line-of-sight component of the SMARP magnetic field maps (blue) and vertical component of the SHARP vector magnetic field maps (green). The SHARP time series shows a minimum between 14-16 July 2010, when the active region passed through the central meridian. This effect appears because the unsigned flux away from the central meridian contains more contribution from the noisier, transverse field. The SMARP time series, on the other hand, shows a maximum at the same time.

To address this issue, we calculated three additional SHARP parameters --- the total line-of-sight unsigned flux, mean gradient of the line-of-sight field, and $R$, listed as the first three parameters listed in Table \ref{tab:keys} under the subheading ``SHARP Data Series'' --- by selecting pixels in the HARP region (e.g. the gold and violet region in Figure \ref{fig:Figure1}h) from line-of-sight magnetic field maps in Heliographic Cylindrical Equal-Area coordinates. We note that the suffix ``\texttt{L}'' in the keyword name refers to the line-of-sight field. Figure \ref{fig:Figure2}b shows the total line-of-sight unsigned flux over time for NOAA Active Region 11087 as calculated from SHARP line-of-sight magnetic field maps (orange). We encourage users to use these three parameters when attempting statistical studies that compare active region summary parameters across both the SMARP and SHARP data sets.

We also calculate uncertainty estimates for the SMARP and SHARP active region summary parameters and plot these as error bars in Figure \ref{fig:Figure2}. Note that the error bars are smaller than the size of the dots for some parameters. To calculate the uncertainties for the SMARP parameters, we assume a homogeneous noise level of 26.4 Mx cm$^{-2}$ per pixel in the MDI line-of-sight magnetic field maps. In fact, Figure 9 of \citet{liu2012} shows that the MDI line-of-sight magnetic field maps contain a spatially-dependent noise distribution ranging from 14.9 Mx cm$^{-2}$ to 55 Mx cm$^{-2}$, with a median value of 26.4 Mx cm$^{-2}$. However, assuming a homogeneous distribution of the highest possible noise, 55 Mx cm$^{-2}$ per pixel, still results in a small relative error. The relative error in the SMARP parameter \texttt{MEANGBL} is $\approx$0.5\% assuming an homogeneously-distributed noise of of 26.4 Mx cm$^{-2}$ per pixel. This increases to $\approx$1\% assuming a noise level of 55 Mx cm$^{-2}$ per pixel. For the SMARP parameters \texttt{USFLUX} and \texttt{R\_VALUE,} the relative errors are $\approx$0.2\% and $\approx$2\%, respectively. This increases to $\approx$0.4\% and $\approx$4\% assuming a noise level of 55 Mx cm$^{-2}$ per pixel. \citet{liu2012} also note that they overestimate noise levels for both the MDI and HMI line-of-sight magnetic field data, since the quiet-Sun data used in their analysis contains some signal from weak-field regions.

We apply the same method to the SHARP parameters derived from line-of-sight magnetic field data. For these data, we assume homogeneous noise level of 6.4 Mx cm$^{-2}$ per pixel in the HMI line-of-sight magnetic field maps as described in Figure 2 of \citet{liu2012}. Section 5 of \citet{bobra14} describes the uncertainty estimates for the SHARP parameters derived from vector magnetic field data (i.e., the last two parameters listed in Table \ref{tab:keys} under the subheading ``SHARP Data Series''). In brief, they used standard deviations and correlation coefficients of the uncertainties from the spectral line inversion code \citep{Centeno2014} to determine the statistical errors of the vector magnetic field. The HMI team computes these statistical errors per observation. They are listed in Table A.4 of \citet{bobra14}.

\begin{figure*}
\plotone{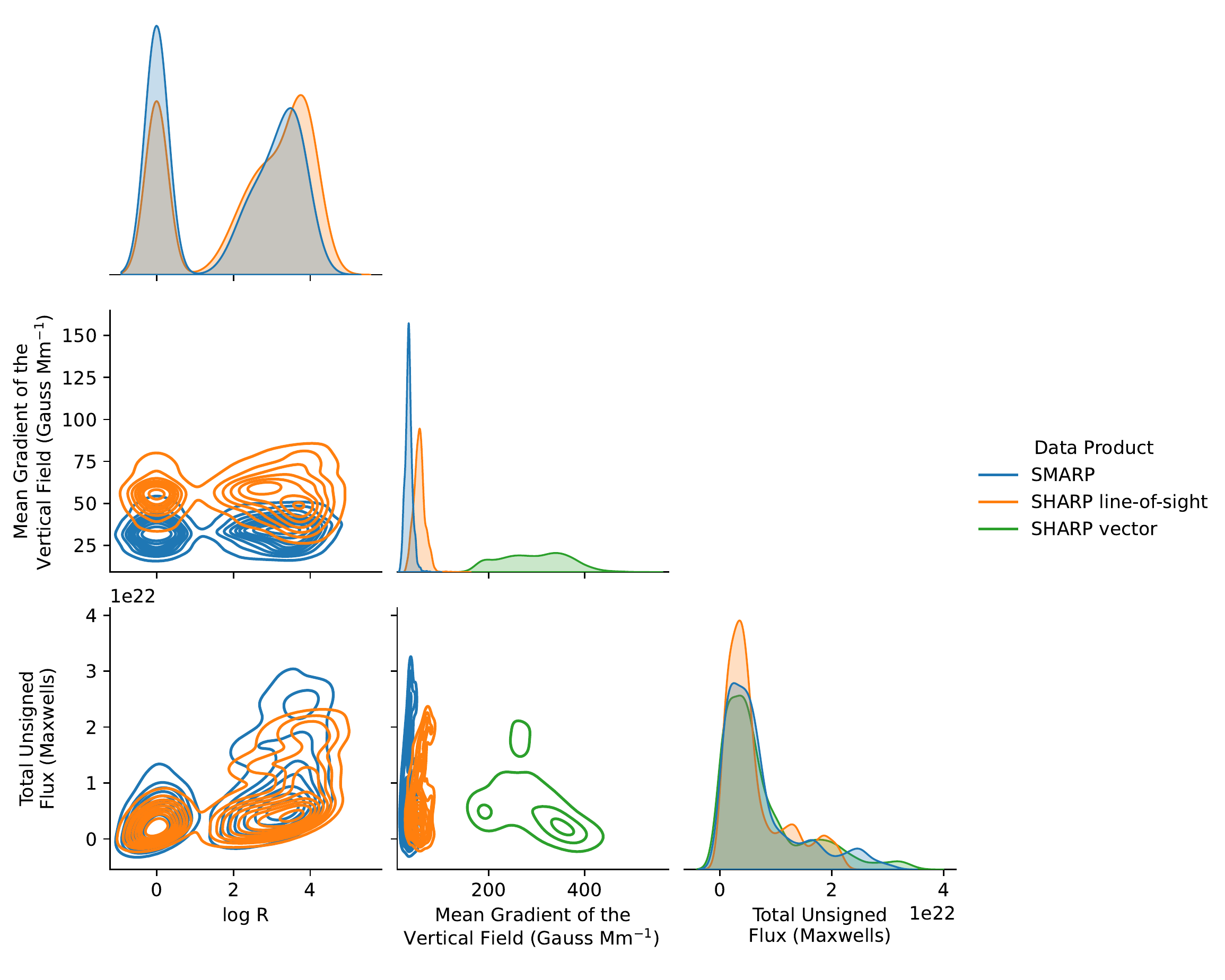}
\vspace{-4mm}
\caption{This figure compares co-temporal values of the $R$, mean gradient of the line-of-sight field, and total line-of-sight unsigned flux parameters described in Table \ref{tab:keys} for 51 co-observed NOAA active regions during the overlap period. We compare three different keyword values --- those calculated from the SMARP line-of-sight magnetic field data, SHARP line-of-sight magnetic field data, SHARP vector magnetic field data. To minimize the effect of projection errors, we only selected keyword values for active regions observed within 70$^\circ$ of central meridian. We placed a further restriction on the quality of the data, encoded in the \texttt{QUALITY} keyword. This leaves 12861 observations, each uniquely describing a NOAA active region at some point in its lifetime. We plot three parameter values per observation, or a total of 38583 values. The diagonal elements of the grid show a kernel density estimate distribution for each of the three parameters. The other elements of the grid show contours of the bivariate distribution between any two parameters.
\label{fig:Figure3}}
\end{figure*}

To illustrate the similarities, and differences, between parameters derived from line-of-sight magnetic field data and those derived from vector magnetic field data, we compared co-temporal values of every parameter listed in Table \ref{tab:keys} for 51 NOAA active regions observed by both HMI and MDI during the overlap period. The diagonal elements of Figure \ref{fig:Figure3} show a kernel density estimate distribution for each of the three summary parameter values. The parameters calculated from the SHARP line-of-sight magnetic field maps show a larger Spearman correlation with the SMARP parameters than those calculated from the SHARP vector magnetic field maps (0.94 and 0.79 for the total line-of-sight unsigned flux and mean gradient of the line-of-sight field, respectively, versus 0.80 and 0.36). The Spearman correlation for the $R$ parameter, which only uses the line-of-sight field, equals 0.88.

\subsection{Limitations of the SMARP Data}\label{subsec:limits}

The SMARP data suffer from a number of problems, and we urge users to proceed with caution. First, optical distortion within the MDI magnetic field maps, mentioned in Section \ref{subsec:maps}, prevents high-precision alignment between the SMARP data and other data sets. Second, as mentioned in Section \ref{subsec:keys}, the MDI line-of-sight magnetic field maps within the SMARP data series contain far more noise \citep[a median error of 26.4 Mx cm$^{-2}$ for the one-minute maps, and 16.2 Mx cm$^{-2}$ for the five-minute maps per Figure 9 of][]{liu2012} than the HMI-line-of-sight magnetic field maps within the SHARP data series \citep[a median error of 6.4 Mx cm$^{-2}$ per Figure 2 of][]{liu2012}. 

Third, MDI only maps the line-of-sight component of the photospheric magnetic field. This only gives users partial information about the magnetic field vector. As mentioned in \ref{subsec:keys}, the line-of-sight component only equals the vertical field at the disk center. It is also known that the sunspot penumbrae contain highly inclined field. This means that the contribution of the vertical and horizontal components of the magnetic field vector to the line-of-sight component changes with viewing angle and within the same active region. This projection effect impacts the time variation of the SMARP active region summary parameters. Users can correct for some of this error using existing methods to approximate the vertical field from the line-of-sight field \citep[e.g.][]{leka2017}.

The MDI mission ended in April 2011. We provide the SMARP data as a legacy product, rather than a gold-standard measurement of the photospheric magnetic field. Before vector magnetic field maps of the photospheric magnetic field were readily available, many studies used the line-of-sight magnetic field to study active region evolution \citep[e.g.][]{Georgoulis2007,mason2010,falconer2011}. By studying both SMARP and SHARP data during the overlap period, we hope users can calibrate legacy observational studies and models that use line-of-sight magnetic field data with modern ones that use vector magnetic field data.

\section{Discussion}\label{sec:discussion}

This paper describes the SMARP data set, which, together with the SHARP data, provide a continuous, seamless set of active region data from Cycle 23 and 24. Combined, the SHARP and SMARP data provide an opportunity to extend statistical studies to a large sample of strong, complex active regions. For example, HMI observed 45 X-class flares to date. The MDI era, on the other hand, witnessed 117 X-class flares.  

Although the SMARP data only include maps of the line-of-sight component of the photospheric magnetic field, whereas the SHARP data include maps of the entire vector, the overlap period provides an opportunity to characterize the differences between the two data sets. Future studies could use this overlap period to develop line-of-sight proxies for the vector magnetic field maps or summary parameters and extend these proxies backward in time. Future studies could also supplement the SMARP and SHARP data with other curated data sets that characterize, for example, subsurface properties of emerging active regions \citep{schunker16}, the surface electric field \citep{Kazachenko14, fisher2020}, and atmospheric phenomena \citep{galvez19}.

At the moment, computational expense and limited storage prevents most users from downloading all the magnetic field maps in the SMARP and SHARP database. Science platforms \citep[e.g.][]{barnes}, or external machines equipped with computing power, software tools, and data sets from multiple solar missions, allow users to rapidly analyze large data sets. These platforms will enable users to derive their own summary parameters from SMARP and SHARP magnetic field maps using methods tailored to the application.

\section{Software and Facilities}

\software{This research used version 3.0.0 \citep{sunpyv300} of the SunPy open source software package \citep{sunpy_community2020}, version 0.6.2 \citep{drms062} of drms \citep{Glogowski2019drms}, version 1.20.2 of NumPy \citep{numpy}, version 0.11.1 \citep{michael_waskom_2020_4379347} of seaborn \citep{seaborn}, version 3.4.2 \citep{matplotlib342} of Matplotlib \citep{matplotlib}, version 1.2.4 \citep{jeff_reback_2021_4681666} of pandas \citep{pandas}, version 1.6.3 \citep{scipy163} of SciPy \citep{scipy}, and version 4.2.1 \citep{thomas_robitaille_2021_4670729} of AstroPy \citep{astropy}. See \citet{monica_g_bobra_smarp} for the code used to reproduce the figures and analysis in this paper.}

\facilities{This research used data from SDO/HMI \citep{schou12} and SoHO/MDI \citep{scherrer95}. These data are publicly available through the JSOC at 
\href{http://jsoc.stanford.edu}{jsoc.stanford.edu}.}

\acknowledgments
NASA grants NNX12AB45G and NNX14AK42G supported the MDI Resident Archive and Stanford Helioseismology Archive, and
grant NNN13D832T (Heliophysics Data Environment Enhancement) supported development of the MDI active region maps.
We would like to thank Jeneen Sommers, Hao Thai, and Arthur B. Amezcua for assisting with the series development.
Part of this research was carried out at the Jet Propulsion Laboratory, California Institute of Technology, 
under a contract with the National Aeronautics and Space Administration (80NM0018D0004). 

\bibliography{bibliography}
\bibliographystyle{aasjournal}
\end{CJK*}
\end{document}